\documentclass[prd,twocolumn,amsmath,amssymb,showpacs]{revtex4}

\usepackage{graphicx}
\usepackage{dcolumn}
\usepackage{bm}
\usepackage{amssymb}

\newcommand{\ignore}[1]{}
\newcommand{\be}{\begin{equation}} \newcommand{\ee}{\end{equation}}
\newcommand{\ba}{\begin{eqnarray}} \newcommand{\ea}{\end{eqnarray}}
\newcommand{\nn}{\nonumber} \renewcommand{\bf}{\textbf}
\newcommand{\ra}{\rightarrow} 
\renewcommand{\d}{\mathrm{d}}

  \newcommand{\epsi}{\varepsilon}

\def\slasha#1{\setbox0=\hbox{$#1$}#1\hskip-\wd0\hbox to\wd0{\hss\sl/\/\hss}}
\def\slashb#1{\setbox0=\hbox{$#1$}#1\hskip-\wd0\dimen0=5pt\advance
       \dimen0 by-\ht0\advance\dimen0 by\dp0\lower0.5\dimen0\hbox
         to\wd0{\hss\sl/\/\hss}}

\begin{document}

\title{The Greisen Equation Explained and Improved} 

\author{Rainer W.~Schiel}

\author{John P.~Ralston}

\affiliation{Department of Physics and Astronomy \\ The University of Kansas \\ Lawrence, Kansas 66045}

\begin{abstract}
Analytic description of the evolution of cosmic ray showers is dominated by the Greisen equation nearly five decades old.  We present an alternative approach with several advantages.  Among the new features are a prediction of the differential distribution, replacing Greisen's form which fails to be positive definite. Explicit comparison with Monte Carlo simulations shows excellent agreement after a few radiation lengths of development. We find a clear connection between Monte Carlo adjustment of Greisen's form and underlying physics, and present a concise derivation with all steps explicit.   We also reconstruct the steps needed to reproduce Greisen's approximate formula, which appears not to have been published previously.  \end{abstract}


\maketitle

\section{Introduction}

The calculation of cosmic ray showers dates back to 1937, when Carlson \& Oppenheimer \cite{carl37} and Bhabha \& Heitler \cite{bhab37} developed the theory of shower evolution. Until computers became powerful enough to run simulations, only analytic methods were available.  The Greisen equation from 1956 \cite{grei56} became the most popular solution.  While state-of-the-art Monte Carlo simulations can yield very accurate results, analytic expressions are still used extensively.  An analytic formula is invaluable for ``back-of-the-envelope'' calculations and order-of-magnitude estimates. Analytic expressions can also be central to simulations:  by combining Monte Carlo calculations at high energies, where ``new physics'' happens, with analytic solutions for the lower energies, computing time can be reduced enormously.

Here we present the Greisen equation and show --- as far as we know --- its derivation for the first time in the literature.   We then introduce an alternative treatment of the physics, the electro-photon model, which allows simple clear treatment without need for the ad-hoc replacements used by Greisen.   We show that the electro-photon ($EP$) equation agrees well with the Greisen equation in its domain of applicability, and works much better where Greisen fails.  Strictly speaking Greisen's formula applies to the integral distribution $\Pi(t, \,E_0 / \epsi_c)$, the number of particles in the shower at depth $t$, given the energy of the incoming particle $E_0$ and the critical energy $\epsi_c$. The corresponding differential distribution $\pi(t, \, E_0, \, E )$ can also be obtained from an equation given by Greisen. $\pi(t, \, E_0, \, E ) \d E$ is the number of particles with energies between $E$ and $E + \d E$. It turns out that for energies close to the energy of the incoming particle, the Greisen differential distribution is negative, as shown in detail below. Even though Greisen did not claim that his equation is valid in the range where the number of particles is small, it is a disturbing fact that his formula has the feature of coming from an unphysical distribution.

Yet Greisen's formula is simple and works amazingly well for many purposes.  Modern Monte Carlo results are invariably presented via formulas marking up Greisen's form with effective parameters or variable substitutions. This suggests there is physics lurking in the empirical re-scalings.  We find it by first deriving an improved formula, valid for both the differential and integral distributions of a shower, in which the electron- and photon- components are merged into one effective degree of freedom.  Naturally this opens up 
the opportunity to tune what is meant by an ``electro-photon'' with numerical work. Meanwhile the improved formula is as simple as Greisen's, while avoiding the unphysical elements and being free of ad-hoc steps. For a cosmic ray with primary energy $E_{0}$, our practical formula is:  
\ba
\pi(t, \, E_0, \, E ) = \frac{A_{EP} e^{-t_{EP}}}{E}  \sqrt{\frac{t_{EP}}{\ln \frac{E_0}{E}}} I_1 \left( 2 \sqrt{t_{EP} \ln \frac{E_0}{E}} \right); \label{res1} \\ 
\Pi(t, \,E_0 / \epsi_c) = A_{EP} e^{-t_{EP}} I_0 \left( 2 \sqrt{t_{EP} \beta_0 } \right). \label{res2}
\ea 
Here $\beta_0 = \ln(\, E_0 / \epsi_c \,) $, and $I_0$,  $I_1$ are modified Bessel functions. The single-species effective depth $t_{EP}$ is given by
\be
t_{EP} = \lambda t + a_{EP} + \beta_0 (1 - \lambda) - 1/2
\ee
where $\lambda = 2 / \sqrt{3} \approx 1.155$, parameters $a_{EP}$ and the normalization $A_{EP}$ are given in Table \ref{tab:params}.  These Monte Carlo-based re-scalings are perfectly consistent with the model's formulation as due to merging two species of particles into one. 

Readers concerned mainly with practical results can compute with Eqs. \ref{res1}, \ref{res2}, which produce rather fine agreement with Monte Carlo simulations.  In the rest of the paper we show how the results were obtained.  Section II recounts the reconstruction of Greisen's steps.  Section III shows how the well-behaved analytic formula is obtained. Section IV compares the electro-photon model to Monte Carlo results, explains the need for re-scalings, and contains Table \ref{tab:params}.

\section{The Greisen Equation}

Greisen's approximation for the number of particles $ \Pi \left( t, \, E_0 / \epsi_c \right)$ in an electromagnetic shower ``in the region where the number of particles is large''\cite{grei56}, long referred to as the ``Greisen equation'', is:
\be \label{eq1956}
\Pi \left( t, \, E_0 / \epsi_c \right) \approx \frac{0.31}{\sqrt{\beta_0}}e^{t \left( 1 - \frac{3}{2}\ln s \right) }.
\ee
Here $t$ is the distance the shower has developed in units of the radiation length, $\beta_0$ is given by
\be
\beta_0 = \ln \frac{E_0}{\epsi_c},
\ee
where $E_0$ is the primary energy, $\epsi_c$ is the critical energy.  The variable $s$ came to be called the {\it shower age}, given by
\be \label{soft}
s = \frac {3 t}{t + 2 \beta_0}.
\ee

\subsection{Derivation of the Greisen Equation}

Unfortunately, Greisen states this equation without giving a derivation.  Most of the  papers and textbooks that use the Greisen equation cite Greisen's 1956 paper, without giving the full derivation either (see, e.g.~\cite{gais90, stan03}). We have reverse-engineered the steps that Greisen took to find his equation and present them in this section.

\subsubsection{Rossi and Greisen, 1941} 
 
A good starting point for the Greisen equation turns out to be the article by Bruno Rossi and Kenneth Greisen \cite{ross41} ($RG$). It summarizes the state of cosmic ray shower physics at that time. The important part for the Greisen equation is the ``Approximation B'' for shower evolution.  Approximation B includes pair production, bremsstrahlung and collision energy losses of the electrons, but neglects the Compton effect. $RG$ start with the asymptotic formulae for Bremsstrahlung and pair production,
\begin{subequations} \label{brempp}
\ba \label{brems}
\frac{\d \sigma}{\d k} & \propto & \frac{1}{k} \left( \frac{4}{3} - \frac{4}{3} y + y^2 \right) \\
\frac{\d \sigma}{\d E} & \propto & \frac{1}{k} \left( 1- \frac{4}{3} x(1-x) \right),
\ea
\end{subequations}
where $k$ is the photon energy, $E$ the electron energy, $y = k / E$ and $x = E / k$. Using Eqs.~\ref{brempp} and an expression for collision energy losses, $RG$ set up the shower evolution equations. They can be solved in a formal sense analytically, using Mellin integrals. We review this shortly: After several pages of calculations, a certain saddle point approximation and dropping a negligible term, the result is
\be \label{eq1941}
\Pi \left( t \right) = \frac{1}{\sqrt{2 \pi} s} \frac{H_1(s) K_1(s,-s)}{\sqrt{ \lambda''_1(s)t+1/s^2 }} \left( \frac{E_0}{\epsi_c} \right)^s \, e^{\lambda_1(s) t},
\ee
where
\be \label{tofs}
t = - \frac{1}{ \lambda'_1 (s) } \left[ \ln \left( \frac{ E_0 }{ \epsi_c } \right) - \frac{1}{s} \right].
\ee
$H_1(s)$, $K_1(s, -s)$ are rather complicated functions that are tabulated in the article and $\lambda_1(s)$ is an analytic function, also given in the article.

\subsubsection{Greisen's approximations}

Several approximations must be done to get from $RG$ to the Greisen equation. The primary idea is to focus on the behavior of the shower around the shower maximum, that is, for shower age parameter $s=1$. 

First, the function $\lambda_1(s)$ is approximated.  A Taylor expansion of $s  \lambda'_1(s)$ around $ s = 1 $ yields:
\ba
s \lambda'_1(s) & \approx & -0.938 + 0.460 (s-1) - 0.037 (s-1)^2 + \nonumber \\ & & + \mathcal{O}\left( (s-1)^3 \right).
\ea
By reasonable guesswork we are able to deduce that Greisen replaced coefficients in the following way:
\begin{eqnarray*}
0.938 & \rightarrow & 1 \\
0.460 & \rightarrow & 1 / 2 \\
0.037 & \rightarrow & 0.
\end{eqnarray*}
This yields:
\be \label{lambda1}
\lambda_1(s) \approx \frac{1}{2} \left( s - 1 - 3 \ln s \right).
\ee

In Eq.~(\ref{tofs}), we can now drop the $1/s$ term (since $\beta_0 \gg 1/s$). Using Eq.~(\ref{lambda1}) we then get
\be \label{soft2}
s \approx \frac{3 t}{t + 2 \beta_0}
\ee
which is just Eq.~(\ref{soft}) used in the Greisen Equation.

With the help of both Eq.~(\ref{lambda1}) and Eq.~(\ref{soft2}) and a little bit of algebra we can rewrite a part of the Rossi and Greisen equation, Eq.~(\ref{eq1941}): 
\be \label{Grexp}
\left( \frac{E_0}{\epsi_c} \right)^s e^{\lambda_1(s) t} = e^{t \left( 1 - \frac{3}{2} \ln s \right)}.
\ee

In the remaining part of Eq.~(\ref{eq1941}), we drop the $ 1/ s^2$ term since it is small compared to $\lambda_1''(s) t$, and evaluate the rest at $ s = 1$:
\be \label{Grconst}
\left. \frac{1}{\sqrt{2 \pi} s} \frac{H_1(s) K_1(s,-s)}{\sqrt{\lambda''_1(s)t}} \right|_{s = 1} = \frac{0.3162}{\sqrt{\beta_0}} \rightarrow \frac{0.31}{\sqrt{\beta_0}}.
\ee

Putting eqs.~(\ref{Grexp}) and (\ref{Grconst}) together, we get the Greisen equation. 

We note that in some places $s$ is evaluated at $s = 1$ whereas at other places it is left as a variable.  The replacements above are not controlled approximations: the Mellin variable $s$ is supposed to be integrated over a certain contour, and there is no way to predict the effects of the substitutions made.  The particular steps above are definitely not designed to work in regions where the distribution deviates from its maximum.  Notice also that the method of Mellin moments applies to both the differential distribution and the integral particle number, but the steps used are not equally good for both.  These facts begin to explain the problem mentioned earlier that the differential distribution corresponding to Greisen's formula is unreliable.  

\subsection{Differential Distribution}

The Greisen equation does not provide a differential distribution, but in Greisen's article \cite{grei56}, we find the following: ``One may compute the approximate number of particles having an energy exceeding E, provided $W_0 \gg E \gg \epsi_0$ {\em ($E_0 \gg E \gg \epsi_c$ in our convention)}  from Eq.~1 {\em (the Greisen equation)} by replacing the coefficient 0.31 with 0.135 and substituting for $\beta_0$ the quantity $\beta = \ln(W_0 / E)$ {\em ($\beta = \ln(E_0/E)$ here)}.'' It might seem that these substitutions come from nowhere and are completely arbitrary, but it can be shown that they follow from Rossi and Greisen, 1941 \cite{ross41}, by just the same steps as the Greisen equation. The new equation reads:
\be
\Pi \left( t, \, E' \geq E \right) = \frac{0.135}{\sqrt{\beta}}e^{t \left( 1 - \frac{3}{2}\ln s \right) }.
\ee

\begin{figure}
\includegraphics[scale=1.0]{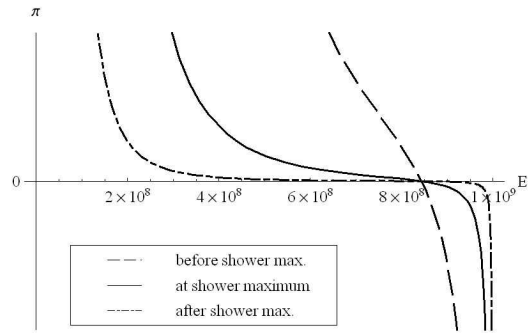}
\caption{Differential distribution as obtained from the Greisen equation before, at and after shower maximum (arbitrary units); energy of the primary is $1 \times 10^9$ .}
\label{fig:greidiffdist} 
\end{figure}

This allows us to calculate the differential distribution $\pi(t, E)$, that is the number of particles with energy $E$ at shower depth $t$: 
\be
\pi(t, \, E) = - \frac{\d \Pi(t, \,E' \geq E)}{\d E}.
\ee
It turns out Greisen's differential distribution is negative for energies close to the energy of the incoming particle, as shown in 
Fig.~\ref{fig:greidiffdist}. Needless to say, this behavior of $\pi(t, \, E)$ is unphysical. Even though the equation is not supposed to be used at energies close to the energy of the incoming particle, this unphysical behavior is a disturbing fact of the replacements and substitutions made.

\subsection{Greisen and Monte Carlo} \label{grvsmc}
  
Comparisons of the Greisen Equation with Monte Carlo calculations are very common. In order to make the Greisen equation agree with the simulations, Fenyves {\it et al.}~\cite{Fenyves:1988gj} introduced two parameters: $a(E)$ shifts the Greisen equation along the radiation length axis, and $A(E)$ re-scales the Greisen equation by a factor. Then 
\be 
\Pi \left( t \right) \approx \frac{0.31 A(E)}{\sqrt{\beta_0}}e^{t_1 \left( 1 - \frac{3}{2}\ln s_1 \right) },
\ee
where $ t_1 = t + a(E) $ and $ s_1 = 3 t_1 / (t_1 + 2 \beta_0) $.

It is impossible to summarize all comparisons of the Greisen equation with the Monte Carlo calculations here. One of the more insightful results was obtained by Sciascio, Piazzoli and Iacovacci ($SPI$) \cite{scia97}. $SPI$ fit $a(E)$ and $A(E)$ for different threshold energies for both the electron component and photon component of electromagnetic showers.  $SPI$'s paper then finds that the data points from the Monte Carlo lie exactly on the lines from the (adjusted) Greisen equation. Given the number of unjustified steps and substitutions this is indeed a puzzle.

\section{The Electro-photon Approach}

In the high energy limit all particles act as if massless and the distinction of electron-positron pairs versus photons ceases to be physically meaningful.  This suggests we should drop the distinction in the mathematics.  The basic assumption for the electro-photon approach is to consider only one species of effective particles, which for discussion we designate the massless {\it electro-photons}.   They will replace the electrons, positrons and photons in the regular shower models.  However the limiting cross sections of pair production and bremsstrahlung are not quite equal.  With one species replacing two, what is meant by the ``radiation length'' must be adjusted to a suitable value.  We gain the freedom to fit the scaling parameter for the effective radiation length after solving the model.

\subsection{Setting up the equations} 

There is little freedom in the cross-section for $EP$s due to dimensional analysis of the high energy limit. Quantum electrodynamics is scale free, so that: 
\be
\frac{\d \sigma}{\d k} \propto \frac{1}{k}.
\ee
This is also an approximation to the Bremsstrahlung cross-section, Eq.~(\ref{brems}), with
\be
\frac{4}{3} - \frac{4}{3} y + y^2 \approx 1.
\ee

Using this cross-section we can write down an evolution equation for the $EP$s,
\be
\frac{\d \pi(t, \, E)}{\d t} = \int_E^\infty \d E' \pi(t, \, E') \frac{1}{E}
\ee
where $t$ is the shower depth in radiation lengths.

This, of course, describes only the gains and not the losses and violates energy conservation. Subtract $\pi(t, \, E)$ on the right hand side to get
\be \label{eveq}
\frac{\d \pi(t, \, E)}{\d t} = \int_E^\infty \d E' \pi(t, \, E') \frac{1}{E} - \pi(t, \, E),
\ee
which incorporates energy conservation. The facts of exact energy conservation will be more easily seen once we have the Mellin transform of this equation. 

\begin{figure}
\includegraphics[scale=1.0]{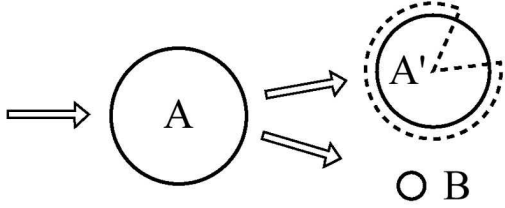}
\caption{Energy conservation in the $EP$-model (the radius represents the energy of the particles): an incoming $EP$ of energy $E$ (A) loses energy by radiating a ``bremsstrahlung'' $EP$ of energy $E'$ (B); A' should now have energy $E-E'$ (solid line), but in the $EP$-model we end up with $1 - E'/E$ electro-photons of energy $E$ (dashed line); obviously, this conserves energy.  }
\label{fig:cartoon} 
\end{figure}

The loss part of Eq.~\ref{eveq} can be pictured as shown in Fig.~\ref{fig:cartoon}: assume one incoming electro-photon of energy $E$. It radiates an $EP$ of energy $E'$. Then the original $EP$ should have only energy $E-E'$ left. However, according to the above equation we end up with $1 - E'/E$ electro-photons with energy $E$. Obviously, this conserves energy. And since in most cases $E'/E \ll 1$ (due to the $1/E$ behavior of the cross-section), this approximation is quite reasonable. 

\subsection{Solution of the electro-photon model}

To solve the evolution equation, Eq.~(\ref{eveq}), we take its Mellin transform (see Appendix \ref{mellin}). The only non-trivial part is the integral on the right hand side:
\ba
\lefteqn{ \int_0^\infty \frac{\d E}{E} E^N \int_E^\infty \d E' \frac{1}{E} \pi( E' ) } \nonumber \\
& = & \int_0^\infty \d E' \pi( E' ) \int_0^{E'} \d E E^{N-2} \nonumber \\
& = & \int_0^\infty \d E' \pi( E' ) \frac{1}{N-1} E'^{N-1} \nonumber \\
& = & \frac{1}{N-1} \pi(N)
\ea
where $\pi(N)$ is the Mellin transform of $\pi (E) $. Therefore the evolution equation is:
\be
\frac{\d \pi(t, \, N)}{\d t} = \left( \frac{1}{N-1} - 1 \right) \pi(t, \, N).
\ee
The $N=2$ mode gives the total energy, and we have
\be
\frac{\d \pi(t, \, 2)}{\d t} = \left( \frac{1}{2-1} - 1 \right) \pi(t, \, 2) = 0.
\ee
Therefore energy is conserved, just as expected.

The evolution equation can be easily solved by an exponential
\be
\pi(t, \, N) = e^{ \left( \frac{1}{N-1} - 1 \right) t} \pi (0, \, N)
\ee
where $\pi(0, \, N)$ describes the incoming particle. We assume one incoming electro-photon with energy $E_0$:
\be \label{initcond}
\pi (t = 0, \, E) = \delta (E-E_0)
\ee
and
\be
\pi(0, \, N) = \int_0^\infty \frac{\d E}{E} E^N \delta (E-E_0) = E_0^{N-1}.
\ee
Therefore, the Mellin transform of the $EP$ distribution is
\be
\pi (t, \, N) = e^{ \left( \frac{1}{N-1} - 1 \right) t} E_0^{N-1}.
\ee

Now we just have to transform this back to $\pi(t, \, E)$. Taking the inverse Mellin transform yields:
\ba
\pi(t, \, E) & = & \frac{1}{2 \pi i} \int_{\alpha -i \infty}^{\alpha +i \infty} \d N E^{-N} \pi(t, \, N) \nonumber \\
& = &  \frac{1}{2 \pi i} \int_{\alpha -i \infty}^{\alpha +i \infty} \d N E^{-N} e^{ \left( \frac{1}{N-1} - 1 \right) t} E_0^{N-1} \nonumber \\
& = & \frac{e^{-t}}{2 \pi i E} \int_{\alpha -i \infty}^{\alpha +i \infty} \d N e^{ \frac{t}{N-1} + (N-1) \ln \frac{ E_0}{ E} }.
\ea

For $\ln ( E_0 / E ) > 0$, we can close the contour in the left half-plane and get
\be
\pi(t, \, E) = \frac{e^{-t}}{E}  \mathfrak{Res} \left( e^{ \frac{t}{N-1} + (N-1) \ln \frac{ E_0}{ E} } \right).
\ee
The residue must be evaluated at the essential singularity $N=1$. It is the residue of an exponential of the form $\exp ( \alpha / x + \beta x)$ which is treated in Appendix \ref{resexp}. Using the results, we get
\be
\pi(t, \, E) = \frac{e^{-t}}{E}  \sqrt{\frac{t}{\ln \frac{E_0}{E}}} I_1 \left( 2 \sqrt{t \ln \frac{E_0}{E}} \right)
\ee
where $I_1$ is a modified Bessel function.

For $\ln ( E_0 / E ) = 0$, that is $ E_0 = E$, we cannot close the contour in the left half-plane \footnote{We thank Roman Buniy for this observation.}. Instead, we go back to the evolution equation, Eq.~\ref{eveq}, and using the initial conditions, Eq.~\ref{initcond}, we see immediately that
\be
\pi(t, \, E=E_0) = e^{-t} \delta(E - E_0).
\ee
Together, we get
\be
\pi(t, \, E) = \left\{ \begin{array}{ll} \frac{e^{-t}}{E} \sqrt{\frac{t}{\ln \frac{E_0}{E}}} I_1 \left( 2 \sqrt{t \ln \frac{E_0}{E}} \right) & \text{for } E < E_0 \\
e^{-t} \delta(E - E_0) & \text{for } E = E_0. \end{array} \right.
\ee
The exponentially decaying $\delta$-function for $E = E_0$ can be understood as the probability for the original particle not to interact down to a certain radiation length. However integrating the $\delta$-function amounts to less than one particle, which compared to the large number of particles in the shower is negligible. Therefore we will drop this term from now on.

\begin{figure}
\includegraphics[scale=1.0]{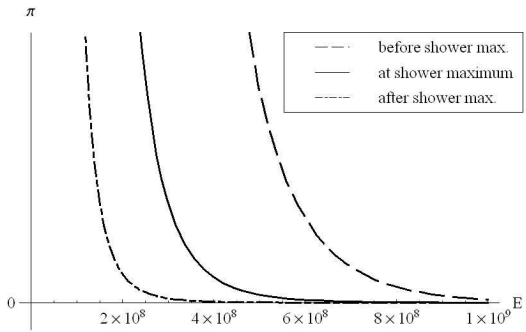}
\caption{Differential distribution for the electro-photon model before, at and after shower maximum (arbitrary units); energy of the primary electro-photon is $1 \times 10^9$.}
\label{fig:epmdiffdist} 
\end{figure}

\begin{figure}
\includegraphics[scale=1.0]{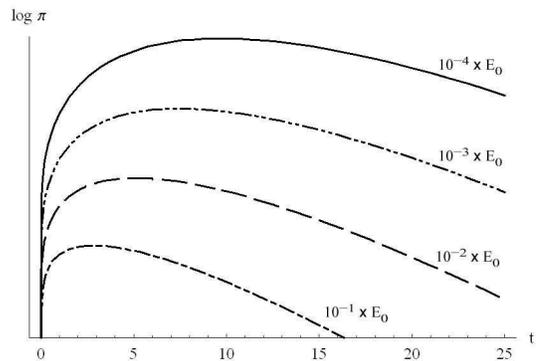}
\caption{Dependence of differential distribution on shower depth in the electro-photon model for certain fractions of the energy of the incoming particle.}
\label{fig:epmdistvst} 
\end{figure}

In Fig.~\ref{fig:epmdiffdist} we show that the differential distribution is everywhere non-negative as expected.  Fig.~\ref{fig:epmdistvst} shows the $t$-evolution for the $EP$ distribution function for certain fractions of the energy of the incoming particle.

\begin{figure}
\includegraphics[scale=1.0]{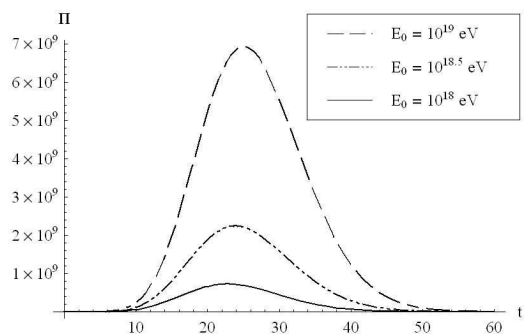}
\caption{Dependence of total particles in the shower on shower depth in the electro-photon model for $E_0 = 10^{19} \text{eV},\, 10^{18.5} \text{eV}$ and $10^{18} \text{eV}$; the critical energy is $\epsi_c = 81 \, \text{MeV}$.}
\label{fig:epmvst}
\end{figure}

To get the total number of electro-photons in the shower, we integrate the differential distribution with respect to $E$. In principle a calculation includes collision losses: particles lose the critical energy $\epsi_c$ per radiation length due to collisions. Therefore, particles with energies less than $\epsi_c$ will drop out of the shower rapidly. Hence, we will only count particles with energies greater than $\epsi_c$:
\be
\Pi(t) = \int_{\epsi_c}^{E_0} \d E \pi(E, t) = e^{-t} I_0 \left( 2 \sqrt{t \ln \frac{E_0}{\epsi_c}} \right).
\ee
Here $I_0$ is a modified Bessel function and $\epsi_c$ is the critical energy. A plot of $\Pi(t) $ for three values of the energy of the incoming $EP$ is shown in Fig.~\ref{fig:epmvst}.

There remains the task of assigning an effective depth $t \ra t_{EP}$ and making use of the numerical results to correctly scale our equations.

\section{Electro-photon model and Monte Carlo Simulations}

We now compare the results from the electro-photon model to Monte Carlo simulations. To do so, three parameters will be fit.  One is an overall scaling factor and the second a shift in $t$-direction. These two steps have been performed previously to make the Greisen equation fit the Monte Carlo simulations, so they need no further justification. 

Earlier we mentioned that the effective cross sections of $EP$s must be a compromise between pair-production and bremsstrahlung processes.  We therefore complete the $EP$ model by adjusting a single effective radiation length, present from the start, as a free parameter.

We want to make use of the existing fit parameters that match the Greisen equation with Monte Carlo results. To do so, we will find the parameters that connect the electro-photon model with the Greisen equation, using analytic methods. Since for most applications the energy of the primary will be much greater than the critical energy, the argument of the Bessel functions will be large and we can use the asymptotic formulae
\ba \label{eq:asympBessel}
I_0 (x) & \approx & \frac{e^x}{\sqrt{2 \pi x}} \left( 1 + \frac{1}{8 x} \right) \nn \\
I_1(x) & \approx & \frac{e^x}{\sqrt{2 \pi x}} \left( 1 - \frac{3}{8 x} \right).
\ea
First, we require that the maximum occurs at the same shower depth. For the Greisen equation, the shower maximum occurs for $t_\text{max} = \beta_0$. Using Eqs.~\ref{eq:asympBessel}, we get $t_\text{max} = \beta_0 - 1/2$ for the $EP$-model. Second, we require that the number of particles be fixed at shower maximum.  For this it is sufficient to keep only the leading term in Eqs.~\ref{eq:asympBessel}. This second requirement leads to the scaling factor of $\Lambda = 0.31 \times \sqrt{4 \pi} \approx 1.099$. Third, we adjust the effective radiation length. This is done by requiring that the second derivatives of the number of particles with respect to shower depth agree at shower maximum. This yields a factor for the effective radiation length of $\lambda = 2 / \sqrt{3} \approx 1.155$. 

Incorporating these results, we get the parametrized $EP$ equations:
\ba
\pi(t, \, E ) = \frac{A_{EP} e^{-t_{EP}}}{E}  \sqrt{\frac{t_{EP}}{\ln \frac{E_0}{E}}} I_1 \left( 2 \sqrt{t_{EP} \ln \frac{E_0}{E}} \right);  \\ 
\Pi(t, \,E_0 / \epsi_c) =   A_{EP} e^{-t_{EP}} I_0 \left( 2 \sqrt{t_{EP} \beta_0 } \right).
\ea 
Here the adjusted shower depth is given by
\be
t_{EP} = \lambda t + a_{EP} + \beta_0 (1 - \lambda) - 1/2.
\ee
$A_{EP}$ and $a_{EP}$ can be obtained from the parameters $A(E)$ and $a(E)$ in the Fenyves {\it et al.}~\cite{Fenyves:1988gj} parametrization of the Greisen equation in the following way:
\ba
A_{EP} & = & \Lambda A(E) \nonumber \\
a_{EP} & = & \lambda a(E) 
\ea
where $\Lambda = 0.31 \times \sqrt{4 \pi} \approx 1.099$ and $\lambda = 2 / \sqrt{3} \approx 1.155$ as above. 

With the numerical data obtained by Sciascio, Piazzoli and Iacovacci \cite{scia97}, we calculate the parameters $A_{EP}$ and $a_{EP}$ for the electro-photon model that agree with these simulations. The parameters are summarized in Table \ref{tab:params}.

Obviously, when fitting new Monte Carlo simulations, it is not necessary to take the detour via the Greisen equation and $A_{EP}$ and $a_{EP}$ can be directly fit from the data.

\begin{figure}
\includegraphics[scale=1.0]{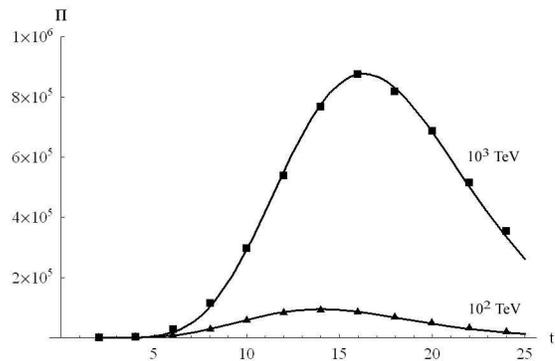}
\caption{Dependence of number of particles in the shower on shower depth for $E_0 = 10^{3}\,  \text{TeV}$ and $10^{2}\,  \text{TeV}$ for electro-photon model (lines) and Monte Carlo \cite{scia97} (squares and triangles). Critical energy $\epsi_c = 81 \, \text{MeV}$; threshold energy $E_{th} = 1 \, \text{MeV}$. }
\label{fig:epmvsgrei} 
\end{figure}

\begin{figure}
\includegraphics[scale=1.0]{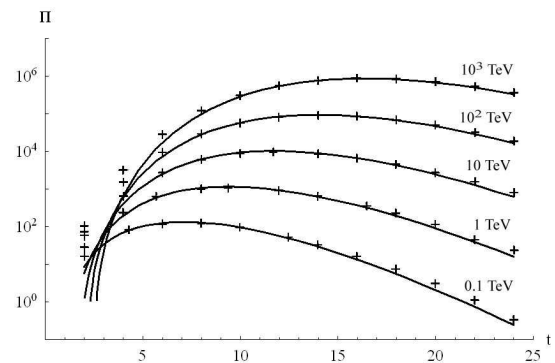}
\caption{Dependence of number of particles in the shower on shower depth for $E_0 = 10^{3}\,  \text{TeV},\, 10^{2}\,  \text{TeV},\, 10\,  \text{TeV},\, 1\,  \text{TeV}$ and $0.1\,  \text{TeV}$ for electro-photon model (lines) and Monte Carlo \cite{scia97} (crosses).  Critical energy $\epsi_c = 81 \, \text{MeV}$; threshold energy $E_{th} = 1 \, \text{MeV}$.}
\label{fig:epmvsgreilog} 
\end{figure}

Incorporating the parameters into the electro-photon model, we get the results cited in the Introduction, and shown in Figures \ref{fig:epmvsgrei} and \ref{fig:epmvsgreilog}. Looking at the plot with the linear scale, Fig.~\ref{fig:epmvsgrei}, there is beautiful agreement between the electro-photon model and the Monte Carlo simulations. The plot with the logarithmic scale, Fig.~\ref{fig:epmvsgreilog}, shows that development over a few radiation lengths is needed to converge to the simulations, just as expected.  The agreement is very good well before the shower maximum and right across the tail of the shower. 

\begin{table}
\caption{$A_{EP}$ and $a_{EP}$ for the electron component and photon component of showers, depending on the threshold energy $E_{th}$. Values are based on data obtained by Sciascio, Piazzoli and Iacovacci \cite{scia97}.}
\label{tab:params}
\begin{ruledtabular}
\begin{tabular}{r|rr|rr}
 & \multicolumn{2}{c|}{electron component} & \multicolumn{2}{c}{photon component} \\
$E_{th}(\text{MeV})$ & $A_{EP}$ & $a_{EP}$ & $A_{EP}$ & $a_{EP}$ \\
\hline
1 & 1.01 & 0.00 & 5.27 & -1.02 \\
5 & 0.82 & 0.22 & 3.27 & -0.80 \\
10 & 0.69 & 0.40 & 2.34 & -0.66 \\
15 & 0.59 & 0.52 & 1.88 & -0.52 \\
20 & 0.55 & 0.61 & 1.59 & -0.42 \\
50 & 0.35 & 0.96 & 0.81 & 0.14 \\
100 & 0.23 & 1.39 & 0.45 & 0.73 \\
\end{tabular}
\end{ruledtabular}
\end{table}

\section{Conclusions}

Greisen's 1956 formula cannot really be derived from coherent mathematical approximations, and it must have been motivated from early numerical work in a time pre-dating high-speed computers.  The underlying physics is extremely simple, and dominated by the $1/k$ dependence of the cross sections, which is a consequence of the scale invariance of quantum electrodynamics.   
A treatment retaining nothing but the leading $1/k$ dependence and combining electrons and photons into one common entity leads to simple analytic formulas with many advantages.  First, the derivation is concise and explicit, and there is no need to replace expressions by proxies. Second, a usable differential distribution is obtained, which is necessary for many purposes.  Finally the results broadly agree with Monte Carlo simulations provided that an effective radiation length is employed, as consistent with the model.  Corrections to this starting point (none are indicated numerically) would probably involve the complicated functions of ratios of two scale parameters seen in the exact solutions and Approximation B. 

\begin{acknowledgments}

We thank Doug McKay, Dave Besson and Roman Buniy for useful discussions and suggestions.  Research supported in part under DOE Grant Number
DE-FG02-04ER14308.  RS partially supported by the University of Kansas General Research Fund, and the KU Energy Research Center.

\end{acknowledgments}

\begin{appendix}

\section{Mellin Transforms} \label{mellin}

\subsection{Definition}

The Mellin transform $f(N)$ of a function $f(E)$ is defined by
\be
f(N) = \int_0^\infty \frac{\d E}{E} E^N f(E).
\ee

\subsection{Inverse Mellin Transform}

The inverse of the Mellin transform is given by
\ba
f(E) & = & \frac{1}{2 \pi i} \int_{\alpha - i \infty }^{\alpha +i \infty } \d N E^{-N} f(N) \nonumber \\
& = & \frac{1}{2 \pi i} \int_C \d N E^{-N} f(N).
\ea
The contour of integration $C$ is parallel to the imaginary axis. 

Proof:
\ba
\lefteqn{ \frac{1}{2 \pi i} \int_C \d N E^{-N} f(N) } \nonumber \\
& = & \frac{1}{2 \pi i} \int_C \d N E^{-N} \int_0^\infty \frac{\d E'}{E'} E'^N f(E') \nonumber \\
& = & \frac{1}{2 \pi i} \int_0^\infty \frac{\d E'}{E'} f(E') \int_C \d N E^{-N} E'^N \nonumber \\ 
& = & \frac{1}{2 \pi i} \int_0^\infty \frac{\d E'}{E'} f(E') \int_C \d N e^{N \ln(E' / E)}.
\ea
We know that
\ba
\lefteqn{ \frac{1}{2 \pi i} \int_{\alpha - i \infty }^{\alpha + i \infty} \d N e^{N \ln(E' / E)} } \nonumber \\
& = & \frac{1}{2 \pi i} e^{\alpha \ln(E' / E) } i \int_{-\infty}^{+\infty} \d x e^{i x \ln(E' / E)} \nonumber \\
& = & e^{\alpha \ln(E' / E)} \delta \left( \ln \frac{E'}{E} \right) = \delta \left( \ln \frac{E'}{E} \right)
\ea
and with this, we have:
\ba
\frac{1}{2 \pi i} \int_C \d N E^{-N} f(N) & = & \int_0^\infty \frac{\d E'}{E'} f(E') \delta \left( \ln \frac{E'}{E} \right) \nonumber \\
& = & f(E).
\ea

\section{Residue of $\exp (\alpha / x + \beta x)$} \label{resexp}

To calculate the residue of 
\be
e^{\frac{\alpha}{x} + \beta x}
\ee
we write the exponential as a series:
\ba
e^{\frac{\alpha}{x} + \beta x} & = & \sum_{i=0}^\infty \frac{1}{i!} \left( \frac{\alpha}{x} + \beta x \right)^i \nonumber \\
& = & \sum_{i=0}^\infty \frac{1}{i!} \sum_{j=0}^i \binom{i}{j} \left( \frac{\alpha}{x} \right)^j \left( \beta x \right)^{i-j}
\ea
For the residue we only need the values of $j$ such that $ -j+(i-j) = -1 $, i.e. $ j = (i + 1)/2 $. So let $i \equiv 2 k + 1 $ which yields $ j = k + 1 $. Then
\ba
\lefteqn{ \mathfrak{Res} \left( e^{\frac{\alpha}{x} + \beta x} \right)} \nonumber \\ 
& = & \mathfrak{Res} \left( \sum_{k=0}^\infty \frac{1}{(2k+1)!} \binom{2k+1}{k+1} \frac{1}{x} \alpha^{k+1} \beta^k \right) \nonumber \\
& = & \sum_{k=0}^\infty \frac{1}{k!\,(k+1)!} \alpha^{k+1} \beta^k = \sqrt{\frac{\alpha}{\beta}} I_1 \left( 2 \sqrt{\alpha \beta} \right)
\ea
where $I_1$ is a modified Bessel function.

\end{appendix}

 \end{document}